\documentstyle[11pt]{article}
\advance \topmargin by -\headheight
\advance \topmargin by -\headsep
     
\evensidemargin \oddsidemargin
\def\rf#1{(\ref{eq:#1})}
\def\lab#1{\label{eq:#1}}
\def\nonu{\nonumber}
\def\br{\begin{eqnarray}}
\def\er{\end{eqnarray}}
\def\be{\begin{equation}}
\def\ee{\end{equation}}
\def\eq{\!\!\!\! &=& \!\!\!\! }
\def\foot#1{\footnotemark\footnotetext{#1}}
\def\lb{\lbrack}
\def\rb{\rbrack}

\def\llb{\left\lbrack}
\def\rrb{\right\rbrack}

\def\lcurl{\left\{}
\def\rcurl{\right\}}
\def\({\left(}
\def\){\right)}

\def\lskip{\vskip\baselineskip\vskip-\parskip\noindent}
\newcommand\partder[2]{{{\partial {#1}}\over{\partial {#2}}}}

\newcommand\Sbr[2]{\Bigl\lbrack\,{#1}\, ,\,{#2}\,\Bigr\rbrack}

\newcommand\pbbr[2]{\lcurl\,{#1}\, ,\,{#2}\,\rcurl} 
\def\a{\alpha}
\def\b{\beta}

\def\h{{1\over 2}}

\def\l{\lambda}

\def\o{\over}

\def\P{\Phi}
\def\pa{\partial}

\def\bpa{{\bar \partial}}

\def\t{\tau}
\def\th{\theta}

\def\wti{\widetilde}
 
\def\cA{{\cal A}}
\def\cB{{\cal B}}

\def\cD{{\cal D}}

\def\cL{{\cal L}}
\def\cM{{\cal M}}

\def\cT{{\cal T}}

\def\cW{{\cal W}}
\def\cX{{\cal X}}

\def\mark{\noindent{\bf Remark.}\quad}

\newtheorem{proposition}{Proposition}

\def\proof{\par{\it Proof}. \ignorespaces} \def\endproof{{$\Box$}\par}

\newcommand\DB{{Darboux-B\"{a}cklund}~}

\def\vp{{\varphi}}

\def\bt{{\bar t}}
\def\pai{\partial^{-1}}
\def\bD{{\bar D}}

\newcommand{\ct}[1]{\cite{#1}}
\newcommand{\bi}[1]{\bibitem{#1}}
\def\Dth{\cD_\theta}
\def\sRes{{\cal R}es}
\def\SKPrm{${\sl SKP}_{{r\o 2},{m\o 2}}$~}
\def\SKPhh{${\sl SKP}_{\h,\h}$~}
\newcommand{\sect}[1]{\setcounter{equation}{0}\section{#1}}

\relax

\newcommand\NPB[3]{{\sl Nucl. Phys.} {\bf B#1} (#2) #3}

\newcommand\CMP[3]{{\sl Commun. Math. Phys.} {\bf #1} (#2) #3}

\newcommand\PLA[3]{{\sl Phys. Lett.} {\bf #1A} (#2) #3}
\newcommand\PLB[3]{{\sl Phys. Lett.} {\bf #1B} (#2) #3}
\newcommand\JMP[3]{{\sl J. Math. Phys.} {\bf #1} (#2) #3}

\newcommand\RMP[3]{{\sl Rev. Mod. Phys.} {\bf #1} (#2) #3}

\newcommand\LMP[3]{{\sl Letters in Math. Phys.} {\bf #1} (#2) #3}
\newcommand\IJMPA[3]{{\sl Int. J. Mod. Phys.} {\bf A#1} (#2) #3}
\newcommand\TMP[3]{{\sl Theor. Mat. Phys.} {\bf #1} (#2) #3}
\newcommand\JPA[3]{{\sl J. Physics} {\bf A#1} (#2) #3}

\newcommand\MPLA[3]{{\sl Mod. Phys. Lett.} {\bf A#1} (#2) #3}

\newcommand\PHSA[3]{{\sl Physica} {\bf A#1} (#2) #3}

\begin{document}
\vspace*{-1.5cm}
\noindent
{\sl solv-int/9801021} \hfill{BGU-98/01/Jan-PH}\\
\phantom{bla}
\hfill{UICHEP-TH/98-1}\\
\begin{center}  
{\Large {\bf Supersymmetric KP Hierarchy: ``Ghost'' Symmetry Structure,
Reductions and Darboux-B\"acklund Solutions}}
\end{center}

\vskip .15in
\begin{center}  
H. Aratyn\foot{E-mail: {\em aratyn@uic.edu}}
\par \vskip .1in \noindent
Department of Physics, University of Illinois at Chicago\\
845 W. Taylor St., Chicago, IL 60607-7059, U.S.A.\\
\par \vskip .15in
E. Nissimov\foot{E-mail: {\em nissimov@inrne.acad.bg}, {\em
emil@bgumail.bgu.ac.il}}  and S. Pacheva\foot{E-mail: 
{\em svetlana@inrne.acad.bg}, {\em svetlana@bgumail.bgu.ac.il}}
\par \vskip .1in \noindent
Institute of Nuclear Research and Nuclear Energy \\
Boul. Tsarigradsko Chausee 72, BG-1784 $\;$Sofia, Bulgaria \\
\vspace{-0.2in}
\begin{center} and \end{center}
\vspace{-0.2in}
Department of Physics, Ben-Gurion University of the Negev \\
Box 653, IL-84105 $\;$Beer Sheva, Israel \\
\end{center}

\begin{abstract} 
This paper studies Manin-Radul supersymmetric KP hierarchy (MR-SKP) in three
related aspects: (i) We find an infinite set of additional (``ghost'')
symmetry flows spanning the same (anti-)commutation algebra as the ordinary
MR-SKP flows; (ii) The latter are used to construct consistent reductions
\SKPrm of the initial unconstrained MR-SKP hierarchy which involves a 
nontrivial modification for the fermionic flows; 
(iii) For the simplest constrained
MR-SKP hierarchy \SKPhh we show that the orbit of \DB
transformations lies on a supersymmetric Toda lattice being a square-root of
the standard one-dimensional Toda lattice, and also we find explicit 
Wronskian-ratio solutions for the super-tau function.
\end{abstract}

\sect{Introduction}

Supersymmetric integrable hierarchies of nonlinear evolution
(``super-soliton'')  equations were
originally proposed \ct{MR-SKP} from purely mathematical motivations, but
soon they attracted active interest also in theoretical physics mainly due
to their close connections with superstring theory \ct{SI-sstring} (for
related studies of supersymmetric integrable systems of Korteveg-de-Vries
or nonlinear-Schr{\"o}dinger type, see \ct{SI-KdV-NLS}).

The scope of the present paper is the supersymmetric Manin-Radul
Kadomtsev-Petviashvili (MR-SKP) hierarchy \ct{MR-SKP} of integrable
super-soliton nonlinear equations within the super-pseudo-differential
operator formulation (see also \ct{SKP-other}; for other formulations see
\ct{SKP-other-1}). We study extensions of MR-SKP hierarchy
incorporating additional (anti-)commuting ``ghost'' symmetries, as well as
reductions of MR-SKP.
We use supersymmetric generalization of several basic concepts in the theory
of integrable systems which up to now have been most actively pursued in
the context of the ordinary (bosonic) KP hierarchy:
Baker-Akhiezer wave functions and tau-functions \ct{Dickey,PvM},
eigenfunctions and squared eigenfunction potentials (see
\ct{noak-addsym,ridge} and references therein).

The advantage of constructing an infinite set of (anti-)commuting ``ghost''
symmetries in the supersymmetric context (see sect.4 below) is two-fold.
On the one hand, it allows us to double the original supersymmetric hierarchy
according to the ``duality'' concept, recently introduced in the context of
the ordinary KP hierarchy \ct{hungry}.
On the other hand, using the ``ghost'' symmetries we are able to define
systematic reductions of the original MR-SKP model to a broad class of
constrained supersymmetric KP hierarchies denoted as
\SKPrm (see eq.\rf{SKP-r-m} below).
These hierarchies possess correct evolution under both even and odd 
isospectral flows. 
The latter turns out to be a non-trivial problem since reductions to
\SKPrm hierarchies are {\em incompatible} with the
original MR-SKP fermionic flows. We provide a solution to this problem by
appropriately modifying MR-SKP fermionic flows while preserving their
original (anti-)commutation algebra, {\sl i.e.}, preserving the integrability
of the constrained \SKPrm systems.

The second part of the paper contains a detailed discussion of
the simplest constrained MR-SKP hierarchy -- \SKPhh (eq.\rf{SKP-h-h} below), 
for which we construct \DB (DB) transformations
preserving both types (even and odd) of the isospectral flows. 
This again is achieved thanks to the above mentioned modification of the
original MR-SKP fermionic flows. Further, we study the pertinent DB-orbit
and discover a new supersymmetric Toda ({\bf s}-Toda) lattice structure
on it. As a consequence of this result we are able to find explicit 
Wronskian-ratio representation for corresponding super tau-function.

Let us mention that
several interesting reduced models of the supersymmetric KP hierarchy
have been previously constructed in the literature in terms of 
super-pseudo-differential operators \ct{SKP-2,SKP-red,AR97,AD97}. 
In particular, the supersymmetric version of AKNS hierarchy  was found which
allows a description in terms of a bosonic \ct{AR97} as well as a fermionic 
\ct{AD97} super-Lax operators.
The various properties and superspace formulation of these models were
worked out, however, their evolution equations involve only even time flows
defining them effectively as reductions of the ${\rm SKP}_2$ hierarchy
\ct{SKP-2}, where only even time flows are present by construction.  

\sect{Background on Manin-Radul Super-KP Hierarchy}
We shall use throughout the super-pseudo-differential calculus \ct{MR-SKP}
with the following notations: $\pa$ and $\cD = \partder{}{\th} + \th \pa$
denote operators, whereas the symbols $\pa_x$ and $\Dth$ will indicate
application of the corresponding operators on superfield functions. As usual,
$(x,\th )$ denote superspace coordinates.
For any super-pseudo-differential operator $\cA = \sum_j a_{j/2} \cD^j$
the subscripts $(\pm )$ denote its purely differential part
($\cA_{+} = \sum_{j \geq 0} a_{j/2} \cD^j$) or its purely
pseudo-differential part ($\cA_{-} = \sum_{j \geq 1} a_{-j/2} \cD^{-j}$),
respectively.
For any $\cA$ the super-residuum is defined as $\sRes \cA = a_{-\h}$.
The rules of conjugation within the
super-pseudo-differential formalism are as follows \ct{AR97}:
$(\cA \cB )^\ast = (-1)^{|A|\, |B|} \cB^\ast \cA^\ast$
for any two elements with gradings $|A|$ and $|B|$;
$\(\pa^k\)^\ast = (-1)^k \pa^k\, ,\,\(\cD^k\)^\ast = (-1)^{k(k+1)/2} \cD^k$
and $u^\ast = u$ for any coefficient superfield.

Finally, in order to avoid confusion we shall also employ the following 
notations: for any super-(pseudo-)\-differential operator $\cA$ and a 
superfield function $f$, the symbol
$\, \cA (f)\,$ will indicate application (action) of $\cA$ on $f$, whereas the
symbol $\cA f$ will denote just operator product of $\cA$ with the zero-order
(multiplication) operator $f$.

MR-SKP hierarchy is defined through the {\em fermionic} Lax operator $\cL$ :
\be
\cL = \cD + f_0 + \sum_{j=1}^\infty b_j \pa^{-j}\cD + 
\sum_{j=1}^\infty f_j \pa^{-j}
\lab{super-Lax}
\ee
expressed in terms of a {\em bosonic} ``dressing'' operator $\cW$ :
\be
\cL = \cW \cD \cW^{-1} \quad ,\quad
\cW = 1 + \sum_{j=1}^\infty \a_j \pa^{-j}\cD + \sum_{j=1}^\infty \b_j \pa^{-j}
\lab{super-dress}
\ee
where $b_j ,\b_j$ are bosonic superfield functions whereas $f_j ,\a_j$ are
fermionic ones and where: 
\be
f_0 = 2\a_1 \quad ,\quad b_1 = - \Dth \a_1 \quad ,\quad 
f_1 = 2\a_2 - \a_1 \Dth \a_1 - 2\a_1 \b_1 - \Dth \b_1
\lab{L-W-rel}
\ee
\mark
The square of MR-SKP Lax operator \rf{super-Lax} is an even operator of the
form:
\be  
\cL^2 = \pa + \Dth b_1 \pai\cD + 
\( 2b_2 + b_1^2 + \Dth f_1 + b_1 \Dth f_0 \) \pai + \ldots
\lab{square-Lax}
\ee
Note that the zero order term in $\cL^2$ vanishes $\Dth f_0 + 2 b_1 = 0\,$ 
due to \rf{L-W-rel}.
\lskip
The Lax evolution eqs. for MR-SKP read \ct{MR-SKP}:
\br
\partder{}{t_l} \cL &=& -\Sbr{\cL^{2l}_{-}}{\cL} = \Sbr{\cL^{2l}_{+}}{\cL}
\lab{super-Lax-even} \\
D_n \cL &=& -\pbbr{\cL^{2n-1}_{-}}{\cL} =
\pbbr{\cL^{2n-1}_{+}}{\cL} - 2\cL^{2n} 
\lab{super-Lax-odd} \\
\partder{}{t_l} \cW &=& - \(\cW \pa^l \cW^{-1}\)_{-} \cW \quad ,\quad
D_n \cW = - \(\cW \cD^{2n-1}\cW^{-1}\)_{-} \cW
\lab{super-dress-eqs}
\er
with the short-hand notations:
\br
D_n = \partder{}{\th_n} - \sum_{k=1}^\infty \th_k \partder{}{t_{n+k-1}}
\quad ,\quad
\pbbr{D_k}{D_l} = - 2 \partder{}{t_{k+l-1}}
\lab{MR-D-n} \\
(t,\th ) \equiv \( t_1 \equiv x ,t_2, \ldots ; \th, \th_1 ,\th_2 ,\ldots \)
\lab{t-th-short}
\er
Accordingly, the super-Zakharov-Shabat (super-ZS) eqs. take the following 
form:
\br
\partder{}{t_k}\cL^{2l}_{+} - \partder{}{t_l}\cL^{2k}_{+} -
\Sbr{\cL^{2k}_{+}}{\cL^{2l}_{+}} = 0
\quad ,\quad
\partder{}{t_k}\cL^{2l-1}_{+} - D_l \cL^{2k}_{+} - 
\Sbr{\cL^{2k}_{+}}{\cL^{2l-1}_{+}} = 0 
\lab{ZS-normal} \\
D_k \cL^{2l-1}_{+} + D_l \cL^{2k-1}_{+} - \pbbr{\cL^{2k-1}_{+}}{\cL^{2l-1}_{+}}
+ 2\cL^{2(k+l-1)}_{+} = 0
\lab{anti-ZS}
\er

\mark  
Let us stress that, unlike the possibility to identify $t_1 \equiv x$ ~(since 
the zero-order term in $\cL^2$ \rf{square-Lax} vanishes), we {\em cannot} 
identify $\th_1 \equiv \th$. Therefore, there is a nontrivial ``evolution''
already with respect to the lowest fermionic flow $D_1$ (which cannot 
in general be identified with $\cD$).
\lskip\par 
The super-Baker-Akhiezer (super-BA) and the adjoint super-BA wave functions
are defined as: 
\be
\psi_{BA} (t,\th ;\l ,\eta ) = \cW \( \psi^{(0)}_{BA} (t,\th ;\l ,\eta )\) 
\quad ,\quad
\psi^{\ast}_{BA} (t,\th ;\l ,\eta ) =
{\cW^\ast}^{-1} \({\psi^\ast}^{(0)}_{BA} (t,\th ;\l ,\eta )\)
\lab{super-BA-def}
\ee
(with $\eta$ being a fermionic ``spectral'' parameter), in terms of the
``free'' super-BA functions:
\br
\psi^{(0)}_{BA} (t,\th ;\l ,\eta ) \equiv e^{\xi (t,\th ;\l ,\eta )}
\quad ,\quad
{\psi^\ast}^{(0)}_{BA} (t,\th ;\l ,\eta ) \equiv e^{-\xi (t,\th ;\l ,\eta )}
\lab{free-super-BA} \\
\xi (t,\th ;\l ,\eta ) = \sum_{l=1}^\infty \l^l t_l + \eta \th + 
(\eta - \l\th )\sum_{n=1}^\infty \l^{n-1} \th_n 
\lab{xi-def}
\er
for which it holds:
\be  
\partder{}{t_k} \psi^{(0)}_{BA} = \pa_x^k \psi^{(0)}_{BA} \quad , \quad
D_n \psi^{(0)}_{BA} = \Dth^{2n-1} \psi^{(0)}_{BA} = 
\pa_x^{n-1} \Dth \psi^{(0)}_{BA} 
\lab{dn-free-super-BA}
\ee
Accordingly, (adjoint) super-BA wave functions satisfy: 
\be
\(\cL^2\)^{(\ast)} \psi^{(\ast)}_{BA} = \pm \l \psi^{(\ast)}_{BA} \quad , \quad
\partder{}{t_l} \psi^{(\ast)}_{BA} = 
\pm \(\cL^{2l}\)^{(\ast)}_{+} (\psi^{(\ast)}_{BA}) \quad ,\quad
D_n \psi^{(\ast)}_{BA} = \pm \(\cL^{2n-1}\)^{(\ast)}_{+} (\psi^{(\ast)}_{BA})
\lab{super-BA-eqs}
\ee
Correspondingly, the defining equations for arbitrary (adjoint-) 
super-eigenfunctions (sEF's) are : 
\be
\partder{}{t_l} \P = \cL^{2l}_{+} (\P) \quad ,\quad
D_n \P = \cL^{2n-1}_{+} (\P) \quad ; \quad
\partder{}{t_l} \Psi = - \(\cL^{2l}\)^{\ast}_{+} (\Psi) \quad ,\quad
D_n \Psi = - \(\cL^{2l}\)^{\ast}_{+} (\Psi) 
\lab{super-EF-eqs}
\ee
with supersymmetric ``spectral'' representations (cf. \ct{ridge}): 
\be
\P (t,\th ) = \int d\l\, d\eta\,\vp (\l,\eta ) \psi_{BA} (t,\th ;\l ,\eta )
\quad ,\quad     \Psi (t,\th ) = 
\int d\l\, d\eta\,\vp^\ast (\l,\eta) \psi^\ast_{BA}(t,\th ;\l ,\eta )
\lab{super-spec}
\ee
For later use let us write down the explicit expression for the ``free'' sEF
$\P^{(0)}$ of the ``free'' $\cL^{(0)} = \cD$. Namely, taking into account
\rf{free-super-BA}--\rf{dn-free-super-BA} and \rf{super-spec}
we get (for definiteness, consider bosonic $\P^{(0)}$):
\br
\partder{}{t_k} \P^{(0)} &=& \pa_x^k \P^{(0)} \quad ,\quad
D_n \P^{(0)} = \Dth^{2n-1} \P^{(0)}
\lab{free-sEF-0} \\
\P^{(0)} (t,\th ) &=& \int d\l\, d\eta\,\,\vp^{(0)}(\l ,\eta)
e^{\xi (t,\th ;\l ,\eta)} \nonu \\
{} &=& 
\int d\l\, \Bigl\lb \Bigl( 1 - \th \sum_{n\geq 1} \l^n \th_n \Bigr)\vp_B (\l)  
+ \Bigl(\th + \sum_{n\geq 1} \l^{n-1} \th_n \Bigr) \vp_F (\l) \Bigr\rb
e^{\sum_{l\geq 1} \l^l t_l}  
\lab{free-sEF}
\er
where $\vp^{(0)}(\l ,\eta) = \vp_F (\l) + \eta \vp_B (\l)$ is arbitrary
``spectral'' density.

The super-tau-function $\t (t,\th )$ is related with the super-residues
of powers of the super-Lax operator \rf{super-Lax} as follows:
\be
\sRes \cL^{2k} = \partder{}{t_k} \Dth \ln \t \quad ,\quad
\sRes \cL^{2k-1} = D_k \Dth \ln \t
\lab{tau-sres}
\ee
Eqs.\rf{tau-sres} follow from the identities:
\br
\partder{}{t_l} \sRes \cL^{2k} = \partder{}{t_k} \sRes \cL^{2l} \quad ,\quad
\partder{}{t_l} \sRes \cL^{2k-1} = D_k \sRes \cL^{2l} \nonu \\
D_l \sRes \cL^{2k-1} + D_k \sRes \cL^{2l-1} + 2 \sRes \cL^{2(k+l-1)} = 0
\lab{sres-id}
\er
which in turn are easily derived from
eqs.\rf{super-Lax-even}--\rf{super-Lax-odd}. In particular, 
for the coefficients of $\cL$ and $\cW$ we have:
\be 
b_1 = \partder{}{t_1} \ln \t \equiv \pa_x \ln \t \quad ,\quad
\a_1 = D_1 \ln \t
\lab{tau-lowest}
\ee

In what follows we shall encounter objects of the form
$\Dth^{-1} (\P \Psi ) = \Dth \pai_x (\P \Psi )$ where $\P ,\Psi$ is a pair of
sEF and adjoint-sEF. Similarly to the purely bosonic case \ct{oevela} one can
show that application of inverse derivative on such products is well-defined
(upto an overall $(t,\th )$-independent constant). Namely, there exists a
unique superfield function -- supersymmetric ``squared eigenfunction
potential'' (super-SEP) $S(\P ,\Psi )$ such that:
$\Dth S(\P ,\Psi ) = \P \Psi$. More precisely the super-SEP satisfies the
relations:
\be
\partder{}{t_k} S(\P ,\Psi ) = \sRes \(\cD^{-1}\Psi\cL^{2k}\P\cD^{-1}\)
\quad ,\quad
D_n S(\P ,\Psi ) = \sRes \(\cD^{-1}\Psi\cL^{2n-1}\P\cD^{-1}\)
\lab{super-Oevel}
\ee
whose consistency follows from the super-ZS eqs.\rf{ZS-normal}--\rf{anti-ZS}.
In particular, eqs.\rf{super-Oevel} for $k=1$ and $n=1$ read:
\be
\pa_x S(\P ,\Psi ) = \sRes \(\cD^{-1}\Psi\cL^2\P\cD^{-1}\) = \Dth (\P \Psi )
= \Dth (\P \Psi )
\quad ,\quad
D_1 S(\P ,\Psi ) = \sRes \(\cD^{-1}\Psi\cL\P\cD^{-1}\) = \P \Psi
\lab{super-Oevel-1}
\ee

\sect{Issue of \DB  Transformations in MR-SKP Hierarchy}
Consider the ``gauge'' transformation of $\cL$ \rf{super-Lax} of the form:
\be
{\wti \cL} = \cT \cL \cT^{-1} \quad ,\quad \cT = \chi \cD \chi^{-1}
\lab{naive-DB}
\ee
which parallels the familiar DB-transformation in the purely bosonic case
\ct{oevela,chau}.  
Requiring the transformed Lax operator ${\wti \cL}$ to obey MR-SKP evolution
eqs. of the same form \rf{super-Lax-even}--\rf{super-Lax-odd} 
as $\cL$ implies that $\cT$ must satisfy: 
\be
\partder{}{t_l}\cT \, \cT^{-1} - \(\cT \cL^{2l}_{+} \cT^{-1}\)_{-} = 0
\quad ,\quad 
D_n\cT \, \cT^{-1} - \(\cT \cL^{2n-1}_{+} \cT^{-1}\)_{-} =
- 2 \({\wti \cL}^{2n-1}\)_{-} 
\lab{DB-consist-0}
\ee

The first eq.\rf{DB-consist-0} is exactly analogous to the purely
bosonic case and implies that $\chi$ must be a sEF \rf{super-EF-eqs} of $\cL$
w.r.t. the even MR-SKP flows. However, there is a problem with the
second eq.\rf{DB-consist-0}. Namely, for the general
(unconstrained) MR-SKP hierarchy it does not have solutions for $\chi$.
In particular, if $\chi$ would be a sEF also w.r.t. fermionic flows
(cf. second eq.\rf{super-EF-eqs}), then the l.h.s. of second
eq.\rf{DB-consist-0} would become zero whereupon we would get the 
contradictory relation: $\({\wti \cL}^{2n-1}\)_{-} = 0$.

Thus, we conclude that the DB-transformations of the general MR-SKP hierarchy
preserve only the bosonic flow equations. In what follows we shall look for
consistent solutions of \rf{DB-consist-0} in the framework of
{\em constrained} MR-SKP systems which will be achieved thanks to a
non-trivial modification of the fermionic MR-SKP flows preserving their
anti-commutation algebra \rf{MR-D-n}.

There is a further essential distinction of DB-transformations for MR-SKP
hierarchy and its purely bosonic counterpart. Calculating
the super-residues of the powers of the DB-transformed Lax operator we
obtain :
\be
\sRes {\wti \cL}^s = \Dth \(\chi^{-1} \cL^s_{+}(\chi )\) +
(-1)^{s+1} \sRes \cL^s
\lab{sres-DB}
\ee

Note the crucial sign factor in front of the second term on the r.h.s. of
eq.\rf{sres-DB}. Together with the first eq.\rf{tau-sres} it implies for the
DB-transformed super-$\t$ function:
\be
{\wti \t} = \chi\, \t^{-1}
\lab{DB-s-tau}
\ee
in contrast with the bosonic case (where we have ${\wti \t} = \chi\, \t$).
\sect{Super-``Ghost'' Symmetries of MR-SKP Hierarchy}
Consider an infinite set $\lcurl \P_{j/2}, \Psi_{j/2}\rcurl_{j=0}^\infty$
of pairs of (adjoint-)sEF's of $\cL$ where those with integer indices are
bosonic, whereas those with half-integer indices are fermionic. Next, let us
introduce the following infinite set of super-pseudo-differential operators:
\be
\cM_{s/2} = \sum_{k=0}^{s-1} \P_{{s-1-k}\o 2} \cD^{-1} \Psi_{k\o 2}
\quad ,\quad   s=1,2,\ldots
\lab{ghost-s}
\ee

which generate an infinite set of flows $\bpa_{s/2}$
($\bpa_{n-\h} \equiv \bD_n \;\; ,\;\; \bpa_k \equiv \partder{}{\bt_k}$) :
\be
\bpa_{s/2} \cW = \cM_{s/2} \cW \quad , \quad 
\bD_n \cL = \pbbr{\cM_{n-\h}}{\cL} \quad ,\quad
\partder{}{\bt_k} \cL = \Sbr{\cM_k}{\cL}
\lab{ghost-flow-Lax}
\ee

On (adjoint-)sEF's entering $\cM_{s/2}$ we allow a {\em non-homogeneous} 
action of the super-flows \rf{ghost-flow-Lax} which parallels the 
construction of generalized ``ghost'' symmetry flows in the bosonic 
case \ct{hungry} (non-homogeneous terms are absent in the traditional 
approach to ``ghost'' symmetry flows \ct{Orlov-etal}) :
\br
\bpa_{s\o 2} \P_{l\o 2} &=& \cM_{s/2} (\P_{l\o 2}) - \P_{{s+l}\o 2} \quad ,\quad
\bpa_{s\o 2} \Psi_{l\o 2} = 
-\cM^{\ast}_{s/2} (\Psi_{l\o 2}) + (-1)^{sl}\Psi_{{s+l}\o 2} 
\lab{ghost-flow-EF} \\
\bpa_{s\o 2} F^{(*)} &=& \pm \cM^{(*)}_{s/2} (F^{(*)})
\lab{ghost-flow-EF-gen}
\er
where $F^{(*)}$ is a generic (adjoint-)sEF not belonging to the set
$\lcurl \P_{j/2},\Psi_{j/2} \rcurl$.

Using \rf{ghost-flow-EF} we arrive at the following:
\begin{proposition}
The infinite set of super-flows $\bpa_{s/2}$ \rf{ghost-s} (anti-)commute both
with the ordinary super-flows of MR-SKP 
\rf{super-Lax-even}--\rf{super-Lax-odd} as well as among themselves:
\br
\Sbr{\partder{}{\bt_s}}{\partder{}{t_l}} = \Sbr{\partder{}{\bt_s}}{D_n} = 0
\quad ,\quad
\Sbr{\bD_s}{\partder{}{t_l}} = \pbbr{\bD_s}{D_n} = 0
\lab{ghost-comm} \\
\Sbr{\partder{}{\bt_s}}{\partder{}{\bt_k}} = 
\Sbr{\partder{}{\bt_s}}{\bD_n} = 0 \quad ,\quad
\pbbr{\bD_i}{\bD_j} = - 2\partder{}{\bt_{i+j-1}}  
\lab{ghost-alg} 
\er
meaning that $\cM_{s\o 2}$ obey the following eqs.:
\br
\partder{}{t_k} \cM_{s\o 2} \eq \Sbr{\cL^{2k}_{+}}{\cM_{s\o 2}}_{-}
\;\; ,\;\,
D_n \cM_k = \Sbr{\cL^{2n-1}_{+}}{\cM_k}_{-} \;\; ,\;\,
D_n \cM_{k-\h} = \pbbr{\cL^{2n-1}_{+}}{\cM_{k-\h}}_{-}
\lab{ghost-comm-1} \\
\partder{}{\bt_k} \cM_l &- &\partder{}{\bt_l} \cM_k - \Sbr{\cM_k}{\cM_l} = 0
\quad ,\quad
\partder{}{\bt_k} \cM_{l-\h} - \bD_l \cM_k - \Sbr{\cM_k}{\cM_{l-\h}} = 0
\lab{ghost-comm-2} \\
\bD_k \cM_{l-\h} &+& \bD_l \cM_{k-\h} - \pbbr{\cM_{k-\h}}{\cM_{l-\h}} =
- 2\cM_{k+l-1}
\lab{ghost-anticomm-2}
\er
\label{proposition:super-comm}
\end{proposition}
In checking eqs.\rf{ghost-comm-1}--\rf{ghost-anticomm-2} we make use of
several useful identities for super-pseudo-differential operators:
\br
\Sbr{\cB_b}{\P_{s\o 2}\cD^{-1}\Psi_{k\o 2}}_{-} &=&
\cB_b (\P_{s\o 2}) \cD^{-1}\Psi_{k\o 2} - 
\P_{s\o 2} \cD^{-1} \cB_b^\ast (\Psi_{k\o 2})
\lab{B-b-X}  \\
\Sbr{\cB_f}{\P_{s\o 2}\cD^{-1}\Psi_{k\o 2}}^{(\pm )}_{-} &=&
\cB_f (\P_{s\o 2}) \cD^{-1}\Psi_{k\o 2} + (-1)^s
\P_{s\o 2} \cD^{-1} \cB_f^\ast (\Psi_{k\o 2})
\lab{B-f-X} \\
\(\P_{s\o 2}\cD^{-1}\Psi_{k\o 2}\) \(\P_{j\o 2}\cD^{-1}\Psi_{l\o 2}\)
&=&\cX_{(s,k)}(\P_{j\o 2}) \cD^{-1} \Psi_{l\o 2} +
(-1)^{k(l+j+1)} \P_{s\o 2}\cD^{-1}\cX_{(j,l)}^\ast (\Psi_{k\o 2})
\lab{X1-X2} \\
\(\P_{j\o 2}\cD^{-1}\Psi_{l\o 2}\)^\ast \eq (-1)^{lj+j+l}
\Psi_{l\o 2}\cD^{-1}\P_{j\o 2} \quad, \quad
\cX_{(s,k)}(\P ) \equiv \P_{s\o 2} \Dth^{-1}\(\Psi_{k\o 2} \P\) 
\lab{conj-X} 
\er
where $\cB_b ,\cB_f$ indicate arbitrary bosonic/fermionic purely differential
super-operators, and ${\Sbr{.}{.}}^{(\pm )}$ denotes commutator or
anticommutator whenever the second element is bosonic/fermionic.

\sect{Constrained MR-SKP Hierarchies}

The super-``ghost''-symmetry flows and the corresponding generating 
operators $\cM_{s\o 2}$ \rf{ghost-s}--\rf{ghost-flow-Lax} can be used to
construct reductions of the full (unconstrained) MR-SKP hierarchy. Namely,
since according to Prop.\ref{proposition:super-comm} the super-``ghost''
flows obey the same algebra \rf{ghost-alg} as the original MR-SKP flows,
we can identify an infinite subset of the latter with a corresponding
infinite subset of the former:
\be
\pa_{\ell {r\o 2}} = - \bpa_{\ell {m\o 2}} \quad ,\;\; \ell =1,2,\ldots \;\; ;
\quad \pa_k \equiv \partder{}{t_k} \; ,\; \pa_{k-\h} \equiv D_k \;\; ;\;\;
\bpa_k \equiv \partder{}{\bt_k} \; ,\; \bpa_{k-\h} \equiv \bD_k
\lab{reduct}
\ee
where $(r,m)$ are some fixed positive integers of equal parity, and retain 
only these flows as Lax evolution flows (this is a supersymmetric extension 
of the usual reduction procedure in the purely bosonic case 
\ct{Oevel-chengs}). Eqs.\rf{reduct} imply the identification 
$\(\cL^{r\ell}\)_{-} = \cM_{\ell{m\o 2}}$ for any $\ell$ and, 
therefore, the corresponding reduced MR-SKP hierarchy denoted as \SKPrm
is described by the following constrained super-Lax operator:
\be
\cL_{({r\o 2},{m\o 2})} = \cL^r_{+} + 
\sum_{j=0}^{m-1} \P_{{m-1-j}\o 2} \cD^{-1} \Psi_{j\o 2}
\lab{SKP-r-m}
\ee

The two simplest constrained MR-SKP Lax operators read:
\br 
\cL_{(\h,\h)} \equiv \cL = \cD + f_0 + \P_0 \cD^{-1} \Psi_0
\lab{SKP-h-h} \\
\cL_{(1,1)} = \pa + \Dth f_0 + 2b_1 + \P_0 \cD^{-1} \Psi_\h +
\P_\h \cD^{-1} \Psi_0
\lab{SKP-1-1}
\er
where $\P_0, \Psi_0$ and $\P_\h, \Psi_\h$ are pairs of bosonic and fermionic
(adjoint-)sEF's w.r.t. the bosonic flows (about the fermionic flows, 
see below).

In what follows we shall consider in some detail the simplest constrained
\SKPhh hierarchy \rf{SKP-h-h}, and henceforth we shall skip the
subscript $(\h,\h)$ of \rf{SKP-h-h} for brevity. 

Using identities \rf{B-b-X}--\rf{X1-X2} we find the identity for any integer
power $N$ (for an analogous formula in the purely bosonic case, see 
\ct{EOR95}) :
\be
\(\cL^N\)_{-} =  
\sum_{j=0}^{N-1}\cL^{N-j-1}(\P_0 )\cD^{-1}{\cL^j}^\ast (\Psi_0)
\lab{SKP-id}
\ee
In particular, for the square of \rf{SKP-h-h} we get:
\be  
\cL^2 = \pa + \cL(\P_0 ) \cD^{-1} \Psi_0 + \P_0 \cD^{-1} \cL^{*}(\Psi_0 )
\lab{square-SKP-h-h}
\ee
where again the zero-order term $\Dth f_0 + 2 \P_0 \Psi_0 = 0$ as a particular
case of \rf{L-W-rel}. 

The constrained MR-SKP Lax operator \rf{SKP-h-h} satisfies consistently the
bosonic flow eqs.\rf{super-Lax-even}.
However, we need to make a
non-trivial modification of the original fermionic flows 
\rf{super-Lax-odd} in order to keep them
compatible with the reduction from the general to the constrained MR-SKP
hierarchy. Indeed, taking the $(-)$ part of eqs.\rf{super-Lax-odd} 
for the constrained $\cL$ \rf{SKP-h-h} and using identity \rf{B-f-X}
together with \rf{SKP-id} we obtain:
\br
\( D_n \P_0 - \cL^{2n-1}_{+}(\P_0)\) \cD^{-1} \Psi_0 -
\P_0 \cD^{-1} \( D_n \Psi_0 + \(\cL^{2n-1}\)^\ast_{+}(\Psi_0)\) = 
- 2\(\cL^{2n}\)_{-}   \nonu \\
= -2 \sum_{j=0}^{2n-1} \cL^{2n-1-j}(\P_0)\cD^{-1}{\cL^j}^\ast (\Psi_0)
\lab{naive-D-n}
\er
which leads to apparent contradiction.

Motivated by our previous work \ct{noak-addsym}\foot{In \ct{noak-addsym}
we solved the problem of incompatibility of the standard
Orlov-Schulman additional non-isospectral symmetry flows \ct{Orlova} with the
reductions of the full bosonic KP hierarchy by appropriately modifying the 
original Orlov-Schulman flows.} 
we arrive at the following important:
\begin{proposition}
There exists the following consistent modification of MR-SKP flows $D_n$
\rf{super-Lax-odd} for constrained \SKPhh hierarchy:
\br
D_n \cL = - \pbbr{\cL^{2n-1}_{-} - X^{(2n-1)}}{\cL} = 
\pbbr{\cL^{2n-1}_{+}}{\cL} + \pbbr{X^{(2n-1)}}{\cL} - 2\cL^{2n}
\lab{odd-flow-new} \\
X^{(2n-1)} \equiv 2 \sum_{l=0}^{n-2} \cL^{2(n-l)-3}(\P_0 ) \cD^{-1}
\(\cL^{2l+1}\)^\ast (\Psi_0)
\lab{X-def} \\
D_n \P_0 = \cL^{2n-1}_{+}(\P_0) - 2 \cL^{2n-1}(\P_0) + X^{(2n-1)}(\P_0)
\lab{P-0-flow-new} \\
D_n \Psi_0 = - \(\cL^{2n-1}\)^\ast_{+}(\Psi_0) +
2\(\cL^{2n-1}\)^\ast (\Psi_0) - \( X^{(2n-1)}\)^\ast (\Psi_0)
\lab{Psi-0-flow-new}
\er
The modified $D_n$ flows obey the same anti-commutation algebra
\rf{MR-D-n} as in the original unconstrained case. 
\label{proposition:modified-odd}
\end{proposition}

In checking the correct anti-commutation algebra for $D_n$ \rf{odd-flow-new}
one has to verify the identities:
\be
D_k X^{(2l-1)} + D_l X^{(2k-1)} - \pbbr{X^{(2k-1)}}{X^{(2l-1)}} -
\pbbr{X^{(2k-1)}}{\cL^{2l-1}}_{-} - \pbbr{X^{(2l-1)}}{\cL^{2k-1}}_{-} = 0
\lab{X-id}
\ee
which in turn follow from the definition of $X^{(2n-1)}$ \rf{X-def} together
with identities \rf{B-b-X}--\rf{conj-X}.  
\lskip
\mark
It is straightforward to generalize Prop.\ref{proposition:modified-odd} for
arbitrary constrained \SKPrm hierarchy \rf{SKP-r-m}. Namely, the modified
fermionic flows have the same form as in \rf{odd-flow-new} where in the
expression for $X^{(2n-1)}$ (cf. \rf{X-def}) one has to sum over all pairs
of (adjoint-)sEF's entering the purely pseudo-differential part of
$\cL_{({r\o 2},{m\o 2})}$ in \rf{SKP-r-m}.
\lskip\par  
Let us now consider DB-transformations on $\cL \equiv \cL_{(\h,\h)}$
\rf{SKP-h-h} preserving its constrained form:
\br
{\wti \cL} = \cT \cL \cT^{-1} = 
\cD + {\wti f}_0 + {\wti \P}_0 \cD^{-1} {\wti \Psi}_0
\quad ,\quad \cT = \P_0 \cD \P_0^{-1}
\lab{DB-L-h-h}  \\
{\wti f}_0 = - f_0 - 2\Dth \ln \P_0 \quad ,\quad
{\wti \P}_0 = \cT\cL (\P_0) = \P_0 \pa_x\ln\P_0 + \P_0\Dth f_0 + \P_0^2 \Psi_0
\quad , \quad   {\wti \Psi}_0 = \P_0^{-1}
\lab{DB-coeff}
\er
We have the following useful identities for DB-transformed quantities:
\be
{\wti \cL}^s ({\wti \P}_0) = \cT\cL^{s+1} (\P_0) \;\; ,\;\;
\({\wti \cL}^{s+1}\)^{\ast} ({\wti \Psi}_0) =
(-1)^{s+1} {\cT^\ast}^{-1} {\cL^s}^\ast (\Psi_0) =
(-1)^s \P_0^{-1} \Dth^{-1} \(\P_0 {\cL^s}^\ast (\Psi_0)\)
\lab{L-P-Psi-id}
\ee
There is a further crucial property of the modified $D_n$ flows
\rf{odd-flow-new}--\rf{X-def} :
\begin{proposition}
The conditions for preserving the fermionic flow 
eqs.\rf{odd-flow-new}--\rf{X-def} by the \DB-transformations on
$\cL \equiv \cL_{\h,\h}$ \rf{SKP-h-h} (cf. second eq.\rf{DB-consist-0}):
\be
D_n\cT \, \cT^{-1} - \(\cT \cL^{2n-1}_{+} \cT^{-1}\)_{-} =
- 2 \({\wti \cL}^{2n-1}\)_{-} + {\wti X}^{(2n-1)} + \cT X^{(2n-1)} \cT^{-1}
\lab{DB-consist}
\ee
where $\cT = \P_0 \cD \P_0^{-1}$ and the ``tilde'' refers to DB-transformed
objects, are now satisfied. 
\label{proposition:DB-consist}
\end{proposition}
The proof of \rf{DB-consist} proceeds by using the modified $D_n$ flow 
definitions \rf{X-def}--\rf{Psi-0-flow-new} together with identities
\rf{B-b-X}--\rf{conj-X} and \rf{L-P-Psi-id}.

\sect{The \DB Orbit of the Constrained MR-SKP Hierarchy}

The recursive expression for the chain of the DB-transformations 
\rf{DB-L-h-h}--\rf{DB-coeff}  
of the constrained \SKPhh hierarchy, starting from the ``free'' initial
$\cL_0 = \cD$, reads (the subscript $k$ indicating the
step of DB-iteration) :  
\br
\cL_{k+1} \eq \cT_k \cL_k \cT_k^{-1} = 
\cD +f_{k+1} + \P_{k+1} \cD^{-1} \Psi_{k+1}   \quad , \quad 
\cT_k = \P_{k} \cD \P_{k}^{-1}
\lab{lkp1}\\
\cL_{1} &=& \cT_0 \cD \cT_0^{-1} = \cD -2 \Dth \ln \P_0 + 
\P_0 \(\pa_x \ln \P_0 \) \cD^{-1} \P_0^{-1}
\lab{tdt}
\er
where:  
\br
f_{k+1} &=&  -2 \Dth \ln \P_{k}- f_{k}   \quad ,\quad
\Psi_{k+1} = \P_{k}^{-1} 
\lab{fkp1}\\
\P_{k+1} &=& \P_{k} \pa_x \ln \P_{k}+\P_{k}\Dth f_{k} + \P_{k}^2 \Psi_{k}
\lab{pkp1}
\er
and where $\P_0$ is a sEF of the initial ``free'' $\cL_0 = \cD$  
satisfying the ``free'' version of eq.\rf{P-0-flow-new}
(no $X^{(2n-1)}$ term). Therefore, its explicit expression is given by 
eq.\rf{free-sEF} with substituting $\th_n \to - \th_n$. Further we have:
\be
\P_1 = \pa_x \P_0 \quad ,\quad \Psi_{1}= \P_0^{-1} \quad ,\quad
f_{1} =-2 \Dth \ln \P_0
\lab{pkp0}
\ee
Note, that from \rf{fkp1}--\rf{pkp1} we find: 
\be
2 \P_{k+1} \Psi_{k+1} + \Dth f_{k+1} =
2 \P_{k} \Psi_{k} + \Dth f_{k} = \ldots = 0
\lab{const-term}   
\ee
which is consistent with the absence of a zero-order term in the square of 
$\cL_{k}$ in \rf{lkp1}.

Eq. \rf{fkp1} can easily be rewritten as follows:
\be
f_{k+1}=  -2 \Dth \sum_{i=0}^k (-1)^{k-i} \ln \P_{i} \lab{fkp1a}
\ee
Recalling identity \rf{const-term} we can alternatively rewrite 
eq. \rf{pkp1} as:
\be
\P_{k+1} = - \h \P_{k} \Dth f_{k+1} = \P_{k} \pa_x \ln \P_{k} -
\P_{k}^2 \Psi_{k} \lab{pkp1a}
\ee
from which we obtain:
\be
\P_{k+1} = \P_{k} \sum_{i=0}^k (-1)^{k-i} \pa_x \ln \P_{i} \lab{pkp1b}
\ee
After making the standard substitution $\P_{k} = e^{\varphi_k}$,
we find from the second eq. in \rf{pkp1a} a {\em new} 
super-Toda ({\bf s}-Toda) lattice equation:
\be
\pa_x \varphi_k = e^{{\varphi_{k+1}}-{\varphi_k} } +
e^{{\varphi_{k}}-{\varphi_{k-1}} }  \lab{stoda}
\ee
Note, that by acting on \rf{stoda} with $\pa_x$ we get:
\be
\pa^2_x \varphi_k = e^{{\varphi_{k+2}}-{\varphi_k} } -
e^{{\varphi_{k}}-{\varphi_{k-2}} }  \lab{btoda}
\ee
which has the form of the ordinary one-dimensional Toda lattice equation
but with a {\em doubled} lattice spacing and, of course, the Toda variables
$\varphi_k = \varphi_k (x,t_2,\ldots ; \th ,\th_1, \ldots)$ are now 
superfields.  
Eq. \rf{stoda} can also be rewritten as:
\be
e^{{\varphi_{k+1}}-{\varphi_k} } = \sum_{i=0}^k (-1)^{k-i} \pa_x \varphi_i
\lab{stodb}
\ee  
or
\be
\varphi_{k+1} = \varphi_{k} + \ln \(\sum_{i=0}^k (-1)^{k-i} \pa_x \varphi_i \)
\lab{stodc}
\ee

We now discuss the Wronskian representation for the sEF's $\P_k$.
The {\bf s}-Toda lattice \rf{stoda} can apparently be thought of as 
the square-root of the standard Toda lattice.
We can use this idea to proceed without any technical calculations.
According to the construction given in \ct{avoda} the EF's $\P_{2n}$ 
associated with even lattice points can be given the usual  
Wronskian expressions with the starting ``point'' $\P_0$.
For the same reason, EF's $\P_{2n+1}$ associated with odd lattice points of  
the {\bf s}-Toda lattice will have the usual Wronskian
expressions with the starting ``point'' $\P_1= \pa_x \P_0 \equiv \P_0^{(1)}~$
\rf{pkp0}.

Generally, for $n=0,1,\ldots $ we find by the above arguments:
\be
\P_{2n} = {W_{n+1}\lb \P_0, \P_0^{(1)},\ldots ,\P_0^{(n)} \rb  \o
W_n \lb \P_0, \P_0^{(1)}, \ldots , \P_0^{(n-1)} \rb } 
\quad ,\quad 
\P_{2n+1} = {W_{n+1}\lb \P_0^{(1)},\P_0^{(2)},\ldots ,\P_0^{(n+1)} \rb  \o
W_n \lb  \P_0^{(1)}, \P_0^{(2)}, \ldots , \P_0^{(n)} \rb } 
\lab{p2n1}
\ee
where $W_k \llb f_1 ,\ldots ,f_k \rrb \equiv
\det {\Bigl\Vert} \pa_x^{i-1} f_j {\Bigr\Vert}$, $i,j=1,\ldots, k$, denotes
standard Wronskian determinant (however, with superfield entries in \rf{p2n1})
and where $\P_0^{(k)} \equiv \pa_x^k \P_0$ with $\P_0$   
as in \rf{free-sEF} (with $\th_n \to -\th_n$).

Using \rf{DB-s-tau} and the above Wronskians expressions \rf{p2n1} 
we find by iteration the super-tau functions obtained by $2n$ recursive
steps of the DB-transformations: 
\be
\t^{(2n)} =
{W_n \llb\P_0^{(1)},\ldots ,\P_0^{(n)}\rrb \o
W_{n+1}\llb\P_0 ,\P_0^{(1)},\ldots ,\P_0^{(n)}\rrb }
\quad , \quad
\t^{(2n+1)} =
{W_{n+1}\llb\P_0 ,\P_0^{(1)},\ldots ,\P_0^{(n)}\rrb \o
W_n \llb\P_0^{(1)},\ldots ,\P_0^{(n)}\rrb }
\lab{tau-match}
\ee

Moreover, since for \rf{SKP-h-h} $\pa_x\ln\t = \P_0 \Psi_0$,
for the $k$-step DB iteration we have 
$\pa_x\ln\t^{(k)} = \frac{\P_k}{\P_{k-1}}$ by taking into account 
\rf{DB-s-tau}. The latter equation together with the relation 
$\t^{(k+1)} = \P_k \t_k^{-1}$ true for any DB-step $k$ (cf. \rf{DB-s-tau})
yields an alternative super-tau-function form of $\bf s$-Toda lattice:
\be
\pa_x\ln\t^{(k)}(t,\th ) = \frac{\t^{(k+1)}(t,\th )}{\t^{(k-1)}(t,\th )}
\lab{s-Toda-tau}
\ee
with the short-hand notation \rf{t-th-short}.

In a subsequent paper we plan to discuss several interesting
issues connected with extending the present results: 
(a) construction of a ``doubled'' MR-SKP
hierarchy by providing a super-Lax formulation for the super-``ghost''
symmetry flows (cf. \rf{ghost-comm}--\rf{ghost-alg}) -- a supersymmetric
extension of the double-KP construction of \ct{hungry}; 
(b) general treatment of
arbitrary constrained \SKPrm hierarchies, including
derivation of more general Wronskian-type solutions for the super-tau
function and elucidating their Berezinian origin; 
(c) obtaining consistent formulation of supersymmetric 
two-dimensional Toda lattice as \DB orbit on the ``doubled'' MR-SKP hierarchy
(similar to the purely bosonic case \ct{hungry}) and of supersymmetric
analogs of random (multi-)matrix models;
(d) study of possible connections of super-tau functions, on one hand, and
partition functions and joint distribution functions in random matrix models
in condensed matter physics (cf. ref.\ct{Yshai-Kamoto}), on the other hand.

{\bf Acknowledgements.}
The authors gratefully acknowledge support by the NSF grant {\sl INT-9724747}.
The work of H.A. was  supported in part by the U.S. Department of Energy
grant {\sl DE-FG02-84ER40173}~ and the work of E.N. and S.P. was supported in
part by Bulgarian NSF grant {\sl Ph-401}.

\end{document}